\documentclass[prl,twocolumn]{revtex4}%
\usepackage{graphicx}
\usepackage{dcolumn}
\usepackage{bm}
\begin{document}

\title {Raman spectra of misoriented bilayer graphene}

\author{P. Poncharal$^{1}$, A. Ayari$^{1}$, T. Michel$^{2,3}$, J. -L.  Sauvajol$^{2}$}

\affiliation{(1)-Laboratoire de Physique de la Mati\`ere Condens\'ee et Nanostructures (UMR CNRS 5586), Universit\'e Claude Bernard, 69622 Villeurbanne, France\\
(2)-Laboratoire des Colloides, Verres et Nanomat\'eriaux (UMR CNRS 5587), Universit\'e Montpellier II, 34095 Montpellier Cedex 5, France\\
(3)-Present address: Laboratoire Pierre Aigrain, Ecole Normale Sup\'erieure, 75 231 Paris, France}

\date{\today}

\begin{abstract}

We compare the main feature of the measured Raman scattering spectra from single layer graphene with a bilayer in which the two layers are
arbitrarily misoriented. The profiles of the 2D bands are very similar having only one component, contrary to the four found for commensurate
Bernal bilayers. These results agree with recent theoretical calculations and point to the similarity of the electronic structures of single
layer graphene and misoriented bilayer graphene. Another new aspect is that the dependance of the 2D frequency on the laser excitation energy
is different in these two latter systems.

\end{abstract}
\vskip 1.5cm

\maketitle

\newpage

%\section{Introduction}

Single-layer graphene (called graphene in the following), defined as a two-dimensional honeycomb lattice of carbon atoms, has rencently attracted major attention from the physics research
community \cite{novoselov1, Zhang1}. Part of the interest lies in the nature of the electronic band structure which permits carriers to behave as massless Dirac fermions with
a vanishing density of states at the Fermi level \cite{Geim1}. These properties are destroyed as soon as two graphene layers are staked in
Bernal AB configuration (referred in the following as Bernal bilayer) as the electronic dispersion curve is no longer linear \cite{Ohta1, latil1}.
The two main processes that are mostly used for the production of graphene are mechanical exfoliation \cite{Novoselov2} and expitaxial
growth on SiC \cite{Berger1}. Epitaxial growth is well adapted to scaling-up and electronic integration, but is controversial because until now several
graphene layers are produced, although on rotational disordered configuration (no Bernal AB stacking). Mechanical exfoliation is a convenient
and inexpensive way to produce graphene but has the main drawback of producing huge amounts of multilayer graphitic pieces. The biggest issue for SiC
epitaxial growth concerns its ability to preserve the linear electronic structure of graphene despite the presence of several
misoriented layers. According to Hass \textit{et al.} \cite{Hass1}, the interesting electronic transport properties are preserved. Recent calculations by Latil \textit{et al.} \cite{Latil2}
seems to comfort these results. The first step in understanding this important system consists in studying two misoriented graphene sheets and comparing them
to graphene.

The main difficulty in this experiment consists in making sure that the observed system is indeed what we seek: i.e. a superposition
of two graphene layers which are arbitrarily misoriented one with the respect to the other (a rotational stacking fault).
The work of Ferrari \textit{et al.} \cite{Ferrari1} has pointed out the importance of Raman spectroscopy in order to identify
unambiguously single layer graphene from Bernal multilayers graphene. However, if the calculation by Latil \textit{et al.} \cite{Latil2} turns out to be correct,
graphene and misoriented bilayer graphene will exhibits identical Raman signature. We need to use an other independant technique to characterize our system.

In this paper, we focus mainly on the study of misoriented bilayer graphene although an example of a graphene misoriented on top of a Bernal bilayer
will be shown. We took advantage of overlapping samples to probe graphene and misoriented bilayer graphene. These systems are characterized with Atomic
Force Microscopy \cite{AFM1} and probed using Raman spectroscopy. We will show that the Raman spectrum of a misoriented bilayer graphene exhibits a single Raman peak, its position depending on the
excitation energy. We will state that the use of Raman spectroscopy to identify unambiguously the nature of multilayer graphene is not as
simple as previously thought.

%\section{Experimental set-up}

Graphene layers were prepared by using mechanical exfoliation of graphite \cite{Novoselov2} and deposited on Si/SiO$_2$ substrate with 290-295 nm thermally
grown oxide (commercially available from IBS \cite{ibs1}). This oxide thickness allows rapid localization of interesting pieces with an optical microscope.
After selecting the most promising pieces, we perform AFM measurements(tapping mode, 300 kHz cantilever with 40 N/m spring constant).

Raman spectra were recorded using two spectrometers. For 488 and
514.5 nm  excitation wavelenghts, we use a Jobin-Yvon T64000
spectrometer operating in triple configuration (1800 gr/mm grating
mode) coupled with a liquid nitrogen cooled CCD camera. For 633 nm
excitation wavelenght, we use a Jobin-Yvon Aramis spectrometer,
(1800 gr/mm grating configuration), with a Pelletier cooled CCD
camera. Excitation laser light was focussed on the substrate using
a confocal microscope with a 1 $\mu$m typical spot size. The 633
nm laser spot on sample was smaller than at 514.5 nm and 488 nm,
due to the different experimental setup. The laser beam power was
set to 3.5 mW on sample for all studied wavelengths. We also
measured Raman spectra between 2 and 6 mW and observed no adverse
heating effects.

%\section{Results}

%\subsection{AFM characterization of the bilayer graphene samples}

Our principal sample consists in two overlapping graphene which
were rotationally disordered. The AFM image of the two graphene
($\alpha$' and $\alpha$'') and their overlap
($\alpha$'+$\alpha$'') is shown in figure 1. We checked it was
indeed two overlapping graphene as follows: the thickness of each
layer relative to the substrate was measured to be 0.7$\pm$0.1 nm.
The edge of the overlapping graphene ($\alpha$'') on top of the
underlying graphene ($\alpha$') was found to be 0.4$\pm$0.1 nm.
Finally, the total thickness ($\alpha$'+$\alpha$'') on top of the
SiO$_2$ was measured at 1.0$\pm$0.1 nm. The color, observed
through optical microscope, also confirms the thickness of the
sheets. Concerning their stacking configuration, we do not know, a
priori, if the overlapped layers will be Bernal or misoriented
(rotationally disordered). However, it is clear that random
overlapping pieces of graphene will not, with a high probability,
be stacked following a Bernal configuration with the lower layer.
Anyway, as the AB-bilayer Raman fingerprint is well known
\cite{Ferrari1, malard1}, a straightforward comparison will show
if the stacking is or is not Bernal like.

%\subsection{Raman data}

Figure 2 shows the Raman spectra for 633 nm excitation wavelength
recorded on the two graphene and compared to the overlap area. We
can immediately see that the spectrum of overlapping layers
strongly differs from a reference Bernal bilayer Raman spectrum.
The unique 2D peak compares well to graphene signature although
its energy dependance is different. These two observations suggest
that the overlapping is indeed a misoriented bilayer graphene.

Interestingly, the width (FWHM) of the 2D band of the misoriented bilayer graphene is smaller (19 cm$^{-1}$) than for
graphene (26 cm$^{-1}$). This result should be compared with turbostratic graphite which also exhibits a single 2D peak, but with a width of
about 40 cm$^{-1}$ \cite{Pimenta1}.

Figure 2 also compares the G and D bands in graphene and misoriented bilayer graphene. For the graphene, the G peak has been
measured at 1585.6 cm$^{-1}$ while it is at 1583 cm$^{-1}$ for the overlapping area (2.6 cm$^{-1}$ shift). The small shift in G peak as a
function of the number of layers has already been mentioned in the literature \cite{gupta1}. Note as well that despite the rotational disorder, the D band is
identical on graphene alone and on the overlapping area.

In the following, we focus on the excitation dependence of the position of the 2D peak in graphene and misoriented bilayer graphene. When the laser wavelength is
reduced to 514.5 nm, the difference between the positions of the two peaks reduces (figure 3, bottom). The small asymmetry
in the 514.5 nm spectrum can be explained rather simply by a small contribution of the single graphene sheet; the laser spot been slightly larger on
the T64000 spectrometer than on the Aramis spectrometer, thus a part of the beam probes the monolayer near by. The up shift is still measurable.
When a laser excitation wavelength of 488 nm is used, the difference between the positions of the two peaks vanishes (Figure 3, top).
For 514.5 nm and 488 nm wavelengths, the graphene signal contribution does not allow accurate determination of the width of the misoriented graphene bilayer 2D peak.
As the overlapping signal plus a graphene contribution gives a narrower peak than graphene alone, it is clear that the peak width reduction is still present.

Table I summarize the results for two misoriented graphene sheets for the three wavelengths used.

%\section{Discussion}

The first conclusion to draw from these measurements is that the interaction between the misoriented graphene layers is weak as it does not split the
electronic dispersion curve because only one component is observed contrary to the case of a Bernal bilayer (four components). This observation is in
agreement with theoretical calculation \cite{Latil2} as well as transport measurement carried out on rotationnally disordered graphene multilayer grown on SiC \cite{Hass1}.
However, the different energy dependence of the 2D band shows that the two systems are not completely equivalent. As the 2D band Raman shift involves
both the electronic band structure and the phonon dispersion curve \cite{thomsen1}, change in either (or both) distribution could induce a shift. In the
framework of an intervalley double-resonance process (DR), the incident photon selects the k vector of the resonant electronic state (E$_{elec}$(k)) in the
vicinity of the K point. The enegy loss E2$_D$ depends on the iTO phonon with wavevector q=2k involved in the DR process as E2$_D$(q) = 2E$_{iTO}$(q) \cite{mafra1}.

In order to explain the misoriented bilayer graphene 2D peak shift
compared to graphene, a first hypothesis could be that the
interaction opens a gap in the electronic band structure. In this
case, we expect the electronic band structure of the misoriented
bilayer graphene, for a given k, to be above the graphene band
structure. Indeed, for an identical excitation energy, the
selected k vector should be smaller due to the gap opening. Thus
we expect a down shift of the 2D band (q been smaller as q = 2k)
contrary to what is observed. Another argument against this first
hypothesis is that we expect a broadening of the 2D band if the
electronic band structure is disturbed (e.g. like for the
Bernal-Bilayer case). In our case, the width of the 2D band does
not increase in the misoriented bilayer compared to graphene. The
unique narrow 2D peak and the shift toward high wave number
compared to graphene leads us to propose a second hypothesis: for
a misoriented bilayer, the weak interaction modifies the phonon
dispersion curve while leaving the electronic band structure
typically unaffected.

Within this proposition, and from our experimental data, we can extract linear fits for the dispersion of the phonon mode. These coefficients are reported in table II.

It was shown that two misoriented graphene layers exhibit, like a single layer, a single peak around 2700 cm$^{-1}$. Is it still possible then to
discriminate a graphene from misoriented bilayer graphene using only Raman spectrometry? The peak frequency shift is
clearly of little use if the excitation wavelength is at 514.5 nm or 488 nm. We suggest that the G to 2D intensity ratio can be used. It is known that, for the first layers, the G peak intensity increases very fast with the number of
layers \cite{Wang1}. For the overlapping configuration studied above, at 633 nm excitation wavelength, we found a ratio of 0.46 for the overlap.
Measurements over five different graphene sheets (including the two that partially overlap) yield a ratio between 0.8 and 1.5. Note however, that the
ratio G/2D is wavelength \cite{Ferrari1} and gate dependent \cite{ferrari2}. Despite these constraints, measuring the G/2D ratio in addition to the unique 2D peak
may be the only technical means to recognize graphene from misoriented bilayer graphene using only Raman spectroscopy.

    The difficulty in identifying graphene by using only Raman spectra is also illustrated in the following example: among our samples, we have a
graphene partially covered by a torn thicker layer (figure 4). We recorded a Raman spectrum (633 nm) on spot (1) and confirmed it was graphene
(single 2D peak with G/2D ratio 0.82). On spot (2), where the thick layer is isolated, we identify unambiguously it as a Bernal bilayer from the
specific features of its 2D band. The color, which depends on the number of layer, whatever their stacking is, also confirms these conclusions.
The overlapping part (spot 3) could then be either a Bernal trilayers (ABA) or a Bernal bilayer plus a misoriented graphene (ABA').

Figure 4 also displays the experimental 2D Raman spectra from spot 3 compared with a known Bernal trilayer recorded at 633 nm.
As the signature is clearly different, we propose it is a graphene misoriented on top of a Bernal bilayer (ABA'). According to the calculation of
Latil \textit{et al.} \cite{Latil2}, a Bernal bilayer plus a rotationally disordered graphene (ABA') should simply exhibit the sum of the spectra from a
bilayer (i.e. a broad band with 4 components) plus the unique peak of graphene. The experimental fingerprint indeed shows several contributions,
a broad base that is compatible with a Bernal bilayer contribution and a single peak which could correspond to shifted graphene peak.
However, their relative intensity is not what is expected for a simple addition of Bernal bilayer and graphene signal.
Note that on a noisy background this 2D band could be mistaken for a graphene. The color observed on optical microscope
will however rule out this type of error.

Finally one can ask if we were able to measure deviation in inter-plane spacing between A-B and non A-B staking. Unfortunately, even with a higher quality
substrate, Atomic Force Microscopy is not accurate enough to measure such deviation, STM analysis will be required.

%\section{Conclusion}

In conclusion the principal point we have shown is that two
misoriented graphene layers still exhibit a single Raman 2D peak
contrarily to a Bernal bilayer. This confirms the weak interaction
between misoriented layers and brings arguments in favor of the
conservation of the linear dispersion of electronic band in
multilayer rotationally disoriented graphene on SiC. The different
dependence in laser excitation energy of the 2D position does
however show that the phonon dispersion curve is modified. Another
important information is that the unicity of a Raman 2D spectra is
not enough to unambiguously identify graphene, the G over 2D ratio
should also be measured.

The author acknowledges Jean-Roch Huntzinger, Sylvain Latil and Luc Henrard for fruitful conversation. We also thank C. Girit for discussion about graphite source.

\newpage

\begin{table}%\begin{center}
\label{table1}
\caption{Dependence on excitation wavelength (nm) of the 2D frequency (cm$^{-1}$) for graphene and misoriented graphene bilayer.\\ }

\begin{tabular}{cccc}
\hline \hline
Laser excitation wavelength (nm)    &   488 & 514.5 & 633 \\
\hline

Graphene (cm$^{-1}$) & 2701 &   2688 &  2641 \\
Misoriented bilayer graphene (cm$^{-1}$) &  2703 &  2694 &  2650 \\
\hline \hline
\end{tabular}
\end{table}

\begin{table}%\begin{center}
\label{table2} \caption{Dispersion relation for iTO phonons fitted
from experimental Raman shift, for excitation between 633 nm and
488 nm, within the hypothesis of identical electronic band
structure for graphene and misoriented bilayer graphene (using
linear electronic dispersion with v$_F$ = 10$^6$ m.s$^{-1}$).\\ }

\begin{tabular}{ccc}
\hline \hline
iTO phonon energy (eV) & Graphene & Misoriented \\
 = a.q+b & & bilayer graphene \\
\hline
a (eV.\AA) &    0.0415 &    0.0372 \\
b (eV) &    0.1514 &    0.1532 \\
\hline \hline
\end{tabular}
\end{table}

\clearpage

\begin{figure}[!ht]
\rotatebox{270}{\includegraphics[%
width=10cm]{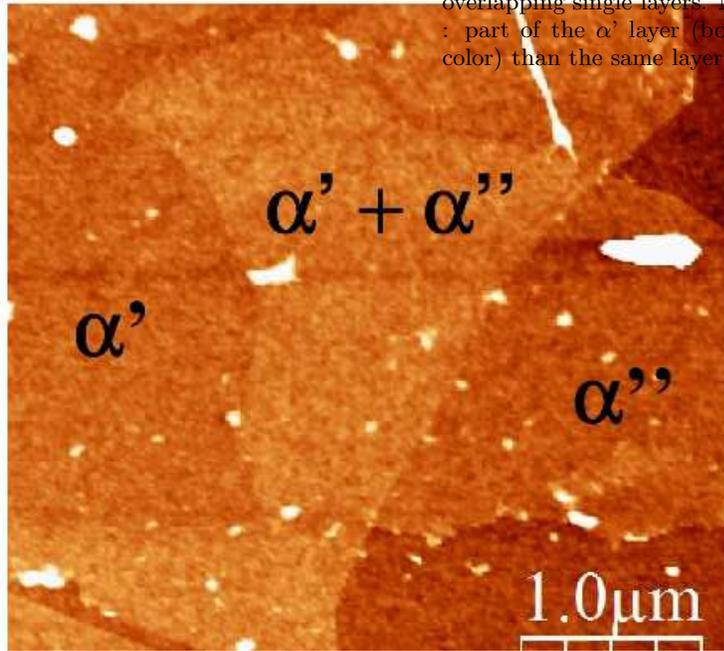}} \vskip 1.5 cm
\caption{(color online) AFM image of two
graphene layers ($\alpha$' and $\alpha$'') and their overlap
($\alpha$'+ $\alpha$''). All measurements (thickness of $\alpha$''
on top of $\alpha$' and thickness of each layers on top of
SiO$_2$) are consistent with the hypothesis of two overlapping
single layers. Note an image "flattening artifact" : part of the
$\alpha$' layer (bottom left) appears higher (lighter color) than
the same layer just above.}
\end{figure}

\begin{figure}[!ht]
\rotatebox{270}{\includegraphics[%
width=10cm]{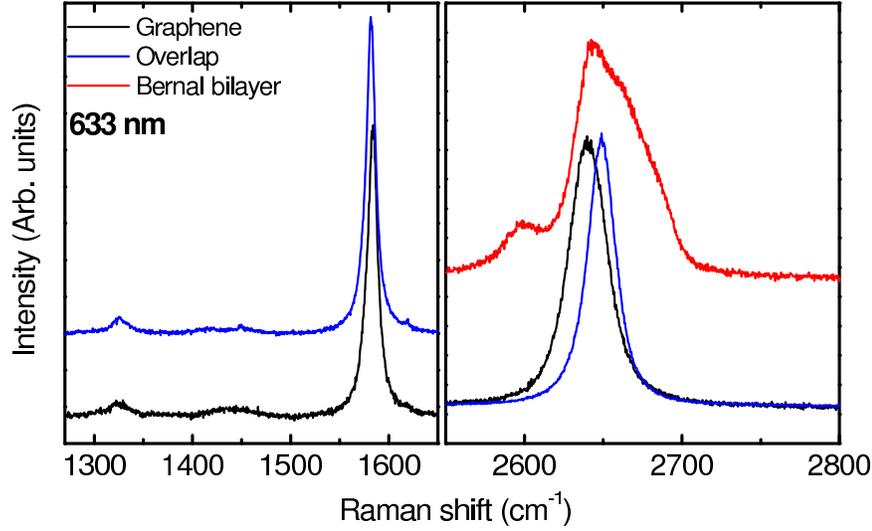}}
\caption{(color online) Raman spectra of
a single graphene sheet ($\alpha$', black lines), Bernal bilayer
(red line) and two overlapping misoriented graphene sheets
($\alpha$' + $\alpha$'', blue lines) at 633 nm. Left : G and D
band range of the graphene and overlapping configuration. Curves
have been vertically offset for clarity and normalized on the G
peak. The frequency shift is about 2.6 cm$^{-1}$ and comparable to
what is reported in the literature. Right : 2D band region for
single graphene sheet ($\alpha$') and overlap ($\alpha$' +
$\alpha$'') compared to Bernal bilayer. The spectrum of $\alpha$''
is almost identical to $\alpha$' and will hardly been seen if
plotted on the same figure. The overlapping graphene spectrum
consists in a single peak clearly shifted compared to single
graphene. It strongly differs form Bernal stacked bilayer (above
curve). Its width (19 cm$^{-1}$) is smaller than the single
graphene peak (26 cm$^{-1}$).}
\end{figure}

\begin{figure}[!ht]
\rotatebox{270}{\includegraphics[%
width=10cm]{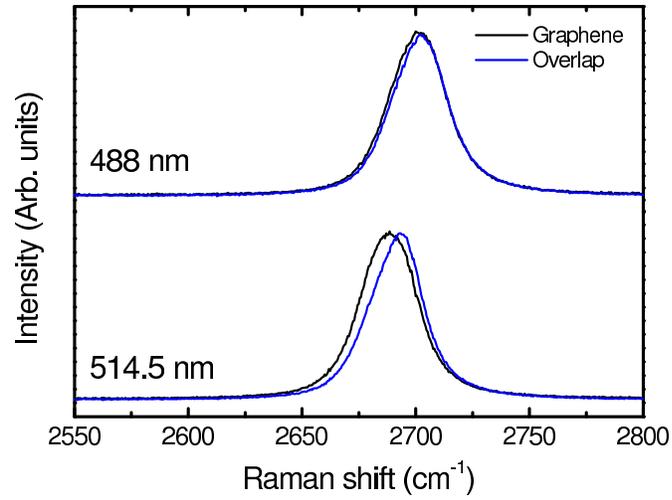}} \caption{(color online) Raman spectrum
of graphene (black lines) and misoriented bilayer graphene (blue
lines) at 488 nm (top) and 514.5 nm (bottom). The difference in
Raman shift is reduced compared to 633 nm. The slight asymmetry is
due to a contribution of the individual graphene sheet (see
text).}
\end{figure}

\begin{figure}[!ht]
\rotatebox{270}{\includegraphics[%
width=10cm]{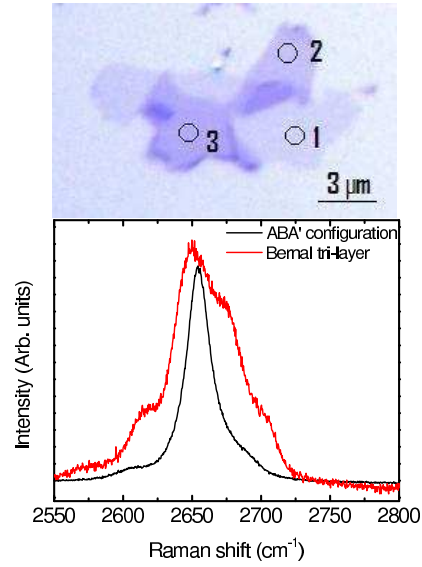}} \caption{(color online) Top : Optical
microscope image of a thin layer (1) covered by a torn thicker
layer (2). Both layers have been independantly probed, (1) has
been found to be graphene and (2) identified as a Bernal bilayer.
The color also confirms these conclusions. The overlapping area
(3) could be ABA (Bernal trilayer) or ABA' (Bernal bilayer plus a
rotationally disordered graphene). Bottom : measured Raman signal
on (3) (black line) compared to a reference Bernal trilayer (red
line) at 633 nm. The clear difference make us state that (3) is a
graphene misoriented on top of a Bernal bilayer.}
\end{figure}

\end{document}